\documentclass[twocolumn]{nature}
\usepackage{graphicx}
\usepackage{epstopdf}
\usepackage[final]{changes}
\usepackage{authblk}
\usepackage{amsmath}
\usepackage{amssymb}
\makeatletter
\let\saved@includegraphics\includegraphics
\AtBeginDocument{\let\includegraphics\saved@includegraphics}
\renewenvironment*{figure}{\@float{figure}}{\end@float}
\makeatother

\newcommand{\RE}{\mathrm{Re}}
\newcommand{\IM}{\mathrm{Im}}
\newcommand{\ket}[1]{|#1\rangle}
\newcommand{\bra}[1]{\langle#1|}

\def\vec{\boldsymbol}
\newcommand{\beq}{\begin{equation}}
\newcommand{\eeq}{\end{equation}}
\newcommand{\beqa}{\begin{eqnarray}}
\newcommand{\eeqa}{\end{eqnarray}}

\begin{document}
\title{Fast spin-valley-based quantum gates in Si with micromagnets}

\author{Peihao Huang$^{1,2}$\thanks{huangph@sustech.edu.cn}\ }
\author{Xuedong Hu$^3$}
\affil{$^1$Shenzhen Institute for Quantum Science and Engineering, Southern University of Science and Technology, Shenzhen 518055, China}
\affil{$^2$Guangdong Provincial Key Laboratory of Quantum Science and Engineering, Southern University of Science and Technology, Shenzhen, 518055, China}
\affil{$^3$Department of Physics, University at Buffalo, SUNY, Buffalo, New York 14260\\}



\date{\today}


\maketitle

\begin{abstract}

An electron spin qubit in silicon quantum dots holds promise for quantum information processing due to the scalability and long coherence. An essential ingredient to recent progress is the employment of micromagnets.  They generate a synthetic spin-orbit coupling (SOC), which allows high-fidelity spin manipulation and strong interaction between an electron spin and cavity photons. To scaled-up quantum computing, multiple technical challenges remain to be overcome, including controlling the valley degree of freedom, which is usually considered detrimental to a spin qubit. Here, we show that it is possible to significantly enhance the electrical manipulation of a spin qubit through the effect of constructive interference and the large spin-valley mixing. To characterize the quality of spin control, we also studied spin dephasing due to charge noise through spin-valley mixing. The competition between the increased control strength and spin dephasing produces two sweet-spots, where the quality factor of the spin qubit can be high. Finally, we reveal that the synthetic SOC leads to distinctive spin relaxation in silicon, which explains recent experiments.

\end{abstract}



\section{Introduction}


A large-scale universal quantum computer can provide enormous computing power for important applications in the future \cite{nielsen2010}. Electron spin qubits in semicondutor quantum dots (QDs) is a possibly scalable system due to device miniaturization and the fabrication technology backed by the semiconductor industry \cite{loss1998, petta2005, hanson_spins_2007, morton_embracing_2011, zwanenburg_silicon_2013}. A spin qubit in a QD can be operated at a temperature above 1 Kelvin, so that more cooling power is available for qubit control, and the common semiconductor substrate makes a quantum device more straightforwardly integrable with classical electronics \cite{yang_operation_2020, petit_universal_2020}. Additionally, a spin qubit in a QD has long relaxation time and long coherence time in isotopically enriched group IV materials (such as silicon and gemanium) \cite{tyryshkin_electron_2012, muhonen2014, veldhorst2015, yoneda2018, zajac_resonantly_2018, watson_programmable_2018, huang_fidelity_2019, hendrickx_fast_2020, hendrickx_four-qubit_2021}, making them ideal hosts for spin qubits.


For a fault-tolerant quantum computer, or the near-term intermediate scale quantum (NISQ) devices, high-fidelity elementary quantum gates is of paramount importance \cite{fowler_surface_2012, preskill_quantum_2018}. Recently, the employment of micromagnets and their associated synthetic spin-orbit coupling (s-SOC) has enabled fast electric dipole spin resonance (EDSR) and strong coupling between a spin qubit and a microwave photon \cite{tokura_coherent_2006, pioro-ladriere_electrically_2008, hu_strong_2012, kawakami_electrical_2014, rancic_electric_2016, yoneda2018, mi_coherent_2018, samkharadze_strong_2018, landig_coherent_2018, benito_electric-field_2019, borjans_resonant_2020, huang_impact_2020}.  However, further improvement to the fidelity of quantum gates for spin qubits in Si QDs could be hindered by the complex environment, particularly the valley degree of freedom in the conduction band and new decoherence channels due to charge noise that are opened by the introduction of micromagnets. 
For example, the valley states lead to a spin-valley hot-spot (SVH) of spin relaxation \cite{yang2013, tahan2014, huang_spin_2014, hao_electron_2014, scarlino_dressed_2017, huang_electrically_2017, huang_spin_2018, borjans_single-spin_2019, struck_low-frequency_2020, zhang_giant_2020}, which could be a detrimental effect.  Charge-noise-induced dephasing and relaxation have also been observed experimentally, though clear theoretical understanding remains lacking \cite{borjans_single-spin_2019, hollmann_large_2020, struck_low-frequency_2020}.  The interplay between s-SOC and valley states remains to be explored \cite{yang2013, huang_spin_2014, huang_electrically_2017, huang_spin_2018, corna_electrically_2018, zhang_giant_2020}. 

In this study, we address the aforementioned problems by studying the spin manipulation, dephasing, and relaxation in a silicon QD in the presence of the valley states, s-SOC, and electrical noise. 
%
We show that, due to an interference effect and the strong spin-valley mixing (SVM), EDSR and spin-photon coupling via the s-SOC can be greatly enhanced.
We have also studied spin pure dephasing due to the $1/f$ charge noise via the SVM, and observed a dephasing hot-spot at the SVH of spin relaxation.
Accounting for both the faster decoherence and manipulation, we find the quality factor for the EDSR (and the spin-photon coupling) peaks on either side of the relaxation hot-spot. Thus, SVM could ultimately benefit rapid high-fidelity quantum gates.
Finally, as a verification of our theory, we explain the experimental signatures of spin relaxation in silicon with a nearby micromagnet at both the high and low magnetic fields. Our results carry clear implications for silicon-based quantum computing, and we hope they stimulate further explorations of valley physics, and interference effects on solid-state qubits.

\section{Results}


\subsection{Model Hamiltonian}

We consider an electron spin qubit in a gated-defined silicon QD in the presence of a micromagnet and an applied magnetic field (Fig.~\ref{schematics} a). The model Hamiltonian is 
\beq
H=H_{\mathrm S} + H_{\mathrm O}+ H_{{\mathrm {SO}}}+  V_{\mathrm e}(\vec{r},t),
\eeq
where $H_{\mathrm S}$ is the bare Hamiltonian of the spin qubit, $H_{\mathrm{O}}= H_{\mathrm V} + H_{\mathrm D}$ is the orbital Hamiltonian consists of the valley term $H_{\mathrm V}$ and the intra-valley orbital term $H_{\mathrm D}$, $H_{\mathrm {SO}}$ is the SOC Hamiltonian, and $V_{\mathrm e}$ is the electric potential from noise or a manipulation field \cite{huang_spin_2014}.
The total magnetic field $\vec{B}(\vec{r})=\vec{B}_0+\vec{B}_1(\vec{r})$ consists of a uniform $\vec{B}_0$ and a position-dependent $\vec{B}_1(\vec{r})$ contribution. The former leads to the bare spin Hamiltonian $H_{\mathrm S}$, $H_{\mathrm S}=\frac{1}{2}E_{\mathrm Z} \vec{\sigma} \cdot \hat{\vec{n}}$, where $E_{\mathrm Z}=g \mu_{\mathrm B}B_0$ is the bare Zeeman splitting, $B_0=|\vec{B}_{\mathrm {MM}} + \vec{B}_{\mathrm {ext}}|$ contains the field $\vec{B}_{\mathrm {MM}}$ from a fully polarized micromagnet and $\vec{B}_{\mathrm {ext}}$ applied externally, $\vec{\sigma}$ is the electron spin operator, and $\hat{\vec{n}}$ is the unit vector along $\vec{B}_0$ assumed to be in-plane. The latter gives rise to a s-SOC, $H_{\mathrm {SO}}=\frac{1}{2}g \mu_{\mathrm B} \vec{\sigma} \cdot \vec{B}_1(\vec{r})$. Without loss of generality, we assume the magnetic field gradient to be in the $x$-direction, such that
\beq
H_{\mathrm {SO}} = \frac{1}{2}g \mu_{\mathrm B} \vec{\sigma}\cdot \vec{b}_{1}x,
\eeq 
where $\vec{b}_{1}\equiv \partial \vec{B}_1/\partial x=[0,0,b_{\mathrm {1t}}]$. 
The s-SOC provides an electric knob to control a spin qubit \cite{tokura_coherent_2006}, while also exposes the qubit to electrical noises. 
{Note that, besides the s-SOC, the intrinsic SOC (i-SOC) is always present in the host material \cite{golovach2004}, and will be included in most of the calculations in this study.}
\subsection{Effective electric dipole of a spin qubit}

%
An important feature of a Si QD is the presence of a low-lying valley excited state, which affects a spin qubit \cite{boykin_valley_2004, friesen_magnetic_2006, goswami_controllable_2007, rahman_engineered_2011, saraiva_intervalley_2011, culcer_valley-based_2012, wu_coherent_2012, yang2013, gamble2013, tahan2014, huang_spin_2014, hao_electron_2014, veldhorst_spin-orbit_2015, boross_control_2016, boross_valley-enhanced_2016, schoenfield_coherent_2017, scarlino_dressed_2017, zimmerman_valley_2017,  huang_electrically_2017, mi_high-resolution_2017, huang_spin_2018, salfi_valley_2018, ferdous_valley_2018, ferdous_interface_2018, ruskov_electron_2018, corna_electrically_2018, penthorn_two-axis_2019, borjans_single-spin_2019, hollmann_large_2020, struck_low-frequency_2020, zhang_giant_2020, zhang_controlling_2021}. In the presence of the s-SOC, the spin and the valley states would mix, making it possible for electrically induced spin flip transitions \cite{tokura_coherent_2006, pioro-ladriere_electrically_2008, hu_strong_2012}. 
{Similarly, the s-SOC also mixes spin and the intra-valley orbital states, leading to electrical field induced spin flip transitions. 
We have shown previously that time-reversal symmetry ($T$-symmetry) plays an important role in the mixing between spin and the intra-valley orbital states \cite{huang_impact_2020}. In particular, the broken $T$-symmetry of the s-SOC modifies the behavior of the spin-orbit mixing and thus the effective magnetic field. However, the previous study relies on perturbation treatment, which is not applicable when the valley splitting is nearly degenerate with the spin splitting. Moreover, in the previous study, the orbital states are assumed to be time-reversal symmetric, which could be violated when the valley states are considered. 
Here, we study the effective dipole due to the mixing of the spin and the valley states non-perturbatively at the degenerate point, and examine explicitly spin properties in the presence of the valley states. 
}

\begin{figure}
\includegraphics[scale=2]{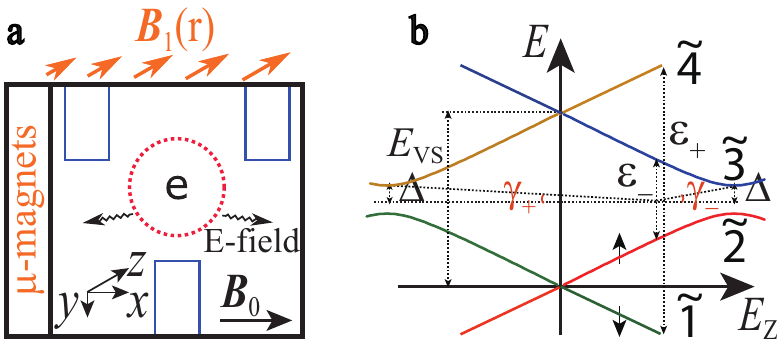}
\caption{Schematics diagrams. (a) Schematic diagram of an electron spin qubit in a gate-defined QD in the presence of micromagnets. An external magnetic field is applied along the $x$-axis and polarizes the micromagnets. A uniform magnetic field $\vec{B}_0$ is along the $x$-axis, and a slanting magnetic field $\vec{B}_1(\vec{r})$ indicated by orange arrows is along the $z$-axis (orthogonal to the $xy$ plane). The slanting field gives rise to a synthetic SOC, which mixes the spin states and valley states. The electric field from phonon or photon leads to spin decoherence or spin manipulation via the synthetic SOC.
(b) The energy level diagram of the mixed spin-valley eigenstates as a function of the Zeeman splitting $E_{\mathrm Z}$. 
The SOC couples the spin-valley product states and results in the eigenstates (denoted as numbers with tildes). 
The mixing angles $\gamma_\pm$ are indicated, where $\tan\gamma_\pm$ is proportional to the splitting $\Delta$ due to SOC and inversely proportional to the energy detuning $\varepsilon_\pm$. 
The broken $T$-symmetry of the synthetic SOC determines the relative phases of the mixings and leads to constructive interference for spin manipulation.
}\label{schematics}
\end{figure}

We denote the spin eigenstates $\ket{\downarrow}$ and $\ket{\uparrow}$, and the two lowest valley eigenstates $\ket{v_0}$ and $\ket{v_1}$ with eigenvalues $\pm E_{\mathrm {VS}}/2$, where $E_{\mathrm {VS}}$ is the valley splitting. The s-SOC mixes $\ket{v_0\uparrow}$ with $\ket{v_1\downarrow}$ \cite{yang2013,tahan2014,huang_spin_2014}, and $\ket{v_0\downarrow}$ with $\ket{v_1\uparrow}$ {[this mixing is omitted in the previous studies \cite{yang2013, tahan2014, huang_spin_2014, zhang_controlling_2021}]}, where the coupling matrix elements are $\Delta_{v_0\uparrow,v_1\downarrow}\equiv \bra{v_0\uparrow} H_{\mathrm {SO}} \ket{v_1\downarrow}$ and $\Delta_{v_0\downarrow,v_1\uparrow}\equiv \bra{v_0\downarrow} H_{\mathrm {SO}} \ket{v_1\uparrow}$. By diagonalizing the coupled Hamiltonian, the spin-valley eigenstates $\ket{\widetilde{1}}$, $\ket{\widetilde{2}}$, $\ket{\widetilde{3}}$, and $\ket{\widetilde{4}}$ are obtained, with the energy spectrum shown in Fig.~\ref{schematics}b.

The transition dipole $r_i^{v_0v_1}=\bra{v_0}r_i\ket{v_1}$ (index $i=x,y,$ or $z$) between the two eigenvalleys is generally non-vanishing {due to disorders at the interface} \cite{yang2013, gamble2013, boross_control_2016, huang_electrically_2017, hosseinkhani_switching_2021}, so that electric field can induce transitions between the spin-valley eigenstates. 
{Note that a detailed calculation of the effect of disorder on the dipole moment $r_i^{v_0v_1}$ would require sophisticated numerical calculations \cite{gamble2013, huang_electrically_2017,  ferdous_interface_2018} 
beyond the scope of this work. As such we treat the dipole matrix element as a phenomenological parameter \cite{yang2013, tahan2014, huang_spin_2014}. 
}

The relevant transition dipole for the spin-flip is $\bra{\widetilde{1}} r_i \ket{\widetilde{2}}$ when $E_{\mathrm Z}<E_{\mathrm {VS}}$, or $\bra{\widetilde{1}}r_i\ket{\widetilde{3}}$ when $E_{\mathrm Z}>E_{\mathrm {VS}}$ \cite{huang_spin_2014}.
The transition dipole takes the form (see Methods)
\beq
\bra{\widetilde{1}}r_i\ket{\widetilde{2}}=-|r_i^{v_0 v_1}| \sin\frac{\gamma_- + \gamma_+}{2}.
\eeq
Here the angles $\gamma_{\mp} =\tan^{-1}(\Delta/\varepsilon_{\mp})$ capture the mixing of $\ket{v_0\uparrow}$ with $\ket{v_1\downarrow}$ and $\ket{v_0\downarrow}$ with $\ket{v_1\uparrow}$, with $\Delta=|\Delta_{v_0\uparrow,v_1\downarrow}|=|\Delta_{v_0\downarrow,v_1\uparrow}| = |g\mu_{\mathrm B} b_{\mathrm {1t}}x^{v_0v_1}/2|$ the amplitude of the spin-valley coupling matrix element, and $\varepsilon_{\mp} = E_{\mathrm {VS}} \mp E_{\mathrm Z}$ the energy detunings as shown in Fig.~\ref{schematics}b.
Similarly, we have $\bra{\widetilde{1}}r_i\ket{\widetilde{3}}=-|r_i^{v_0 v_1}| \cos\frac{\gamma_- + \gamma_+}{2}$.
We emphasize that the "$+$" sign between $\gamma_-$ and $\gamma_+$ in the dipole moments arises from a constructive interference between the two SVM paths, as discussed below.
In comparison, for the case of the i-SOC that presents in the host material, the "$+$" sign is replaced by a "$-$" sign, corresponding to a destructive interference between those mixing paths.

In the limit $\Delta \ll \varepsilon_+ \equiv E_{\mathrm {VS}}+E_{\mathrm Z}$, i.e. weak spin-valley coupling as compared with valley-orbit coupling and/or Zeeman splitting, which is usually satisfied in Si QDs, the transition dipole of the spin qubit for any value of $E_{\mathrm Z}$ can be written as $r_{s}= -|\vec{r}^{v_0v_1}| \eta_{\mathrm {SV}}$, where
\beq
\eta_{\mathrm {SV}}\approx sgn(\varepsilon_-)\sqrt{\frac{1-C_{\mathrm s}}{2}} + \frac{\Delta}{2\varepsilon_+}\sqrt{\frac{1+C_{\mathrm s}}{2}}, \label{etaSV}
\eeq
with $C_{\mathrm s}=\left[1+\frac{\Delta^2}{\varepsilon_-^2}\right]^{-1/2}$ and $sgn(\varepsilon_-)$ the sign of $\varepsilon_-$. The "$+$" sign in front of the second term in $\eta_{\mathrm {SV}}$ is again from a constructive interference between the two SVM paths.
Note the result of $\eta_{\mathrm {SV}}$ is valid both at and away the relaxation hot-spot.
When $|\varepsilon_-|\approx \Delta \ll |\varepsilon_+|$, we can recover the previous results \cite{yang2013, huang_spin_2014}.
When $|\varepsilon_-| \gg \Delta $, we have $\eta_{\mathrm {SV}}= \frac{E_{\mathrm {VS}}\Delta}{E_{\mathrm {VS}}^2 - E_{\mathrm Z}^2}$ (for i-SOC, $\eta_{\mathrm {SV,i-SOC}}= \frac{E_{Z}\Delta}{E_{\mathrm {VS}}^2 - E_{\mathrm Z}^2}$), consistent with perturbative results.

The sign difference in the results for s-SOC and i-SOC is due to the different relative phase between the matrix elements $\Delta_{v_0\uparrow,v_1\downarrow}$ and $ \Delta_{v_0\downarrow,v_1\uparrow}^*$ (see Methods), which are in turn determined by the property of the SOC under time-reversal operation $\Theta$.  In particular, s-SOC breaks the $T$-symmetry, $\Theta H_{\mathrm {s-SOC}} \Theta^{-1}=-H_{\mathrm {s-SOC}}$, so that
$\Delta_{v_0\uparrow,v_1\downarrow} = \bra{\Theta (v_1 \downarrow) } \Theta H_{\mathrm {SO}} \Theta^{-1} \ket{\Theta (v_0\uparrow)} = \Delta_{v_0\downarrow,v_1\uparrow}^*$.  On the other hand, i-SOC conserves the $T$-symmetry, so that $\Delta_{v_0\uparrow,v_1\downarrow} = - \Delta_{v_0\downarrow,v_1\uparrow}^*$.  In short, the breaking of $T$-symmetry by s-SOC modifies the relative phase of the matrix elements $\Delta_{v_0\uparrow,v_1\downarrow}$ (for the mixing between $|\widetilde{2}\rangle$ and $|\widetilde{3}\rangle$) and $ \Delta_{v_0\downarrow,v_1\uparrow}^*$ (for the mixing between $|\widetilde{1}\rangle$ and $|\widetilde{4}\rangle$), thus substantially modifies the properties of the spin qubit due to the interference between the two mixing paths.
{We emphasize again that the observation here extends the previous result on the intra-valley SOM, by considering the valley states and going to the non-perturbative regime. }


Having obtained the effective dipole moment, we explore consequences of the SVM on spin manipulation, spin pure dephasing, and spin relaxation, {and compare with results due to the intra-valley SOM}.

\subsection{Enhanced EDSR and spin-photon coupling}

EDSR via the SOM has been widely used in experiments for fast spin manipulation \cite{golovach_electric-dipole-induced_2006, tokura_coherent_2006, pioro-ladriere_electrically_2008, borhani_spin_2012, nowack_coherent_2007, nadj-perge_spin-orbit_2010,schroer_field_2011, kawakami_electrical_2014, yoneda2018}. 
{In silicon, s-SOC induces both the intra-valley SOM, and SVM. 
When an oscillating electric field of magnitude $\vec{E}_{0}\cos(\omega_{\mathrm Z} t)$ is applied, where $\omega_{\mathrm Z}=g\mu_{\mathrm B} B_0/\hbar$ is the Larmor frequency of an electron spin, then, the Rabi frequency $\Omega_{\mathrm R}$ of the EDSR due to the SVM is}
\beq
\Omega_{\mathrm R} (\vec{E}_0) = e|\vec{E}_{0}\cdot \vec{r}^{v_0v_1}| \eta_{\mathrm {SV}}/\hbar.
\eeq
where $\eta_{\mathrm {SV}}$ is given by Eq.~(\ref{etaSV}). Moreover, the spin-photon coupling $g_{\mathrm s}=\Omega_{\mathrm R}(\vec{E}_{\mathrm {zpf}})$ can also be evaluated, if the EDSR is driven by the electric field $\vec{E}_{\mathrm {zpf}}$ from vacuum fluctuation of a superconducting resonator \cite{blais_cavity_2004, hu_strong_2012, mi_coherent_2018, samkharadze_strong_2018,borjans_resonant_2020}.
The Rabi frequency is thus enhanced via the constructive interference between the SVM paths as compared with the i-SOC induced EDSR.


{As mentioned above, EDSR can arise from both SVM and from the intra-valley SOM. We will evaluate the magnitude of the EDSR due to SVM, and compare the two channels.
We also note here that, in a device with a micromagnet, the i-SOC is also always present, and can contribute to EDSR as well [and also to spin relaxation and dephasing]. Furthermore, the i-SOC can have significant impact on the spin splitting in a QD by modifying the $g$-factor \cite{ferdous_valley_2018}. 
However, in the current study, EDSR and spin relaxation are mainly dominated by the s-SOC [at least in the devices we consider], as evidenced by the faster spin relaxation in the experiment in the presence of micromagnets \cite{borjans_single-spin_2019}. Thus, we neglect the contribution from the i-SOC when the micromagnets are present. 
}

\begin{figure}
\includegraphics[scale=0.6]{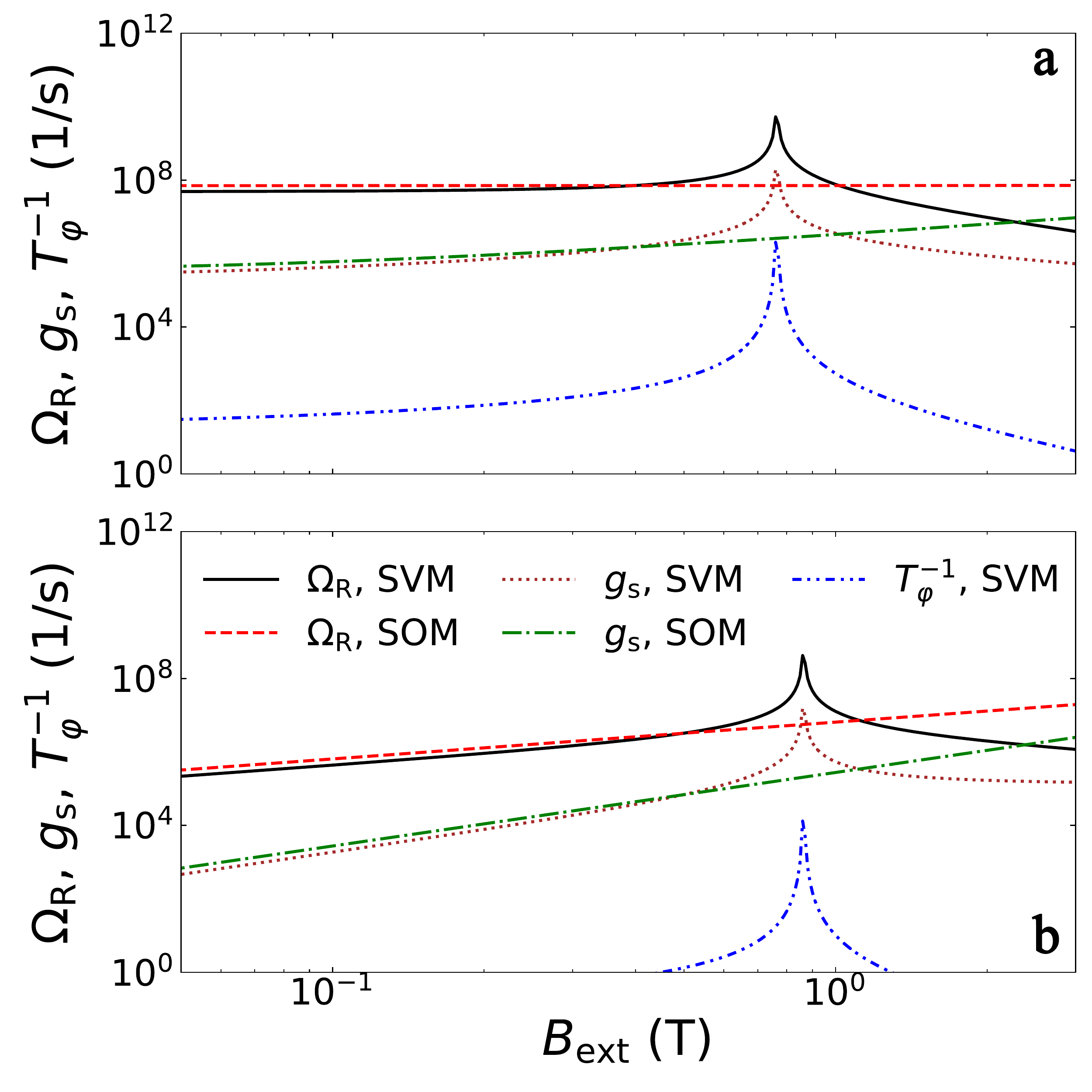}
\caption{\added{Rabi frequencies and spin pure dephasing.} Rabi frequency $\Omega_{\mathrm R}$ of the EDSR, spin-photon coupling $g_{\mathrm s}$ (i.e. vacuum Rabi frequency), and spin pure dephasing $1/T_\varphi$ as a function of the magnetic field $B_0$ {due to the SVM or the intra-valley SOM} for a spin in a silicon QD with the s-SOC (Fig. \ref{fig_spinEDSR} a) or the i-SOC (Fig. \ref{fig_spinEDSR} b). 
Both $\Omega_{\mathrm R}$ and $g_{\mathrm s}$ are greatly enhanced near the SVH. Away from the SVH, 
$\Omega_{\mathrm R}$ and $g_{\mathrm s}$ have weaker $B_0$ dependence for the s-SOC compared to the i-SOC due to the broken $T$-symmetry of the s-SOC. $1/T_\varphi$ is also enhanced near the SVH and has similar $B_0$ dependence for the s-SOC and the i-SOC.
}\label{fig_spinEDSR}
\end{figure}

Fig. \ref{fig_spinEDSR} shows the Rabi frequency $\Omega_{\mathrm R}$ and vacuum Rabi frequency $g_{\mathrm s}$ as a function of the magnetic field $B_0$ due to the s-SOC (Fig. \ref{fig_spinEDSR} a) or the i-SOC (Fig. \ref{fig_spinEDSR} b) {for a device with or without a micromagnet, and using typical quantum dot parameters. Both the SVM and the intra-valley SOM are considered.}  At low magnetic field when $E_{\mathrm Z}\ll E_{\mathrm {VS}}$, the Rabi frequency $\Omega_{\mathrm R}$ via the s-SOC induced SVM stays constant, and the vacuum Rabi frequency $g_{\mathrm s}$ grows linearly with $B_0$ (the cavity frequency is assumed resonance with the spin Lamor frequency, thus the photon energy grows with $B_0$). At $B_0=0.1$ T, $\Omega_{\mathrm R} \sim 10^8$ s$^{-1}$ while $g_{\mathrm s} \sim 10^6$ s$^{-1}$. In comparison, for the i-SOC, $\Omega_{\mathrm R}$ and $g_{\mathrm s}$ shows linear $B_0$ and $B_0^2$ dependence, respectively, and at 0.1 T, $\Omega_{\mathrm R}\sim 10^5$ s$^{-1}$ and $g_{\mathrm s}=10^3$ s$^{-1}$. 
Rabi frequency $\Omega_{\mathrm R}$ via the s-SOC induced SVM has a large magnitude and saturates at low magnetic fields because of the constructive interference attributed to the broken $T$-symmetry of the s-SOC.
As $B_0$ increases, $\Omega_{\mathrm R}$ and $g_{\mathrm s}$ via the s- or i-SOC rise by orders of magnitude near the SVH.
As the magnetic field $B_0$ further increases past the hot-spot, $\Omega_{\mathrm R}$ and $g_{\mathrm s}$ due to the SVM is reduced due to the reduced mixing, while the intra-valley SOM gradually becomes the dominant mechanism for EDSR or spin-photon coupling.
Therefore, the constructive interference and large spin-valley mixing can substantially increase the Rabi frequency of the EDSR and the spin-photon coupling.

\subsection{Spin pure dephasing due to $1/f$ charge noise}

With SVM, pure dephasing for the spin qubit arises at the second order of s-SOC. The effective magnetic noise contributing to spin dephasing due to the SVM is (see Methods) 
\beq
n_{\mathrm {eff}}=eV_{1/f}(r_{\mathrm {dip}}/l_0)\sin\frac{\gamma+\gamma^\prime}{2}\sin\frac{\gamma-\gamma^\prime}{2},
\eeq
where $V_{1/f}$ is the voltage fluctuation from the $1/f$ charge noise, and $r_{\mathrm {dip}}=r^{v_{0}v_{0}}-r^{v_{1}v_{1}}$ is the dipole moment of the valley states.
For two states $\ket{\alpha}$ and $\ket{\beta}$ of interest,
the system dephases as $\exp[{-\phi(\tau)}]$, and $\phi(\tau)= \int_{\omega_{\mathrm c}}^{\infty} d\omega J_{zz}(\omega) [2\sin(\omega \tau/2)/\omega]^2$
 \cite{duan_reducing_1998, taylor_dephasing_2006},
\beqa
J_{zz}(\omega)&=& \frac{2}{\hbar^2}\int_{-\infty}^{\infty}\langle n_{\mathrm {eff}}(0)n_{\mathrm {eff}}(\tau) \rangle \cos(\omega\tau)d\tau,
\eeqa
where 
$n_{\mathrm {eff}}$ is the effective noise obtained above, $J_{zz}(\omega)$ is the spectral density for the noise, and the cutoff frequency $\omega_{\mathrm c}\approx 1$ s$^{-1}$ represents the inverse of the measurement time of coherence dynamics. By evaluating the spin dephasing dynamics according to the equations, the spin pure dephasing rate $1/T_\varphi$ can be obtained \cite{hu2006, huang_spinDephasing_2018}.

Fig. \ref{fig_spinEDSR} also shows the spin pure dephasing rate $1/T_\varphi$ as a function of the magnetic field $B_0$ due to the $1/f$ charge noise via the s- or i-SOC induced SVM. {[The contribution of the intra-valley SOM to spin pure dephasing is negligible since the intra-valley orbital splitting is far off-resonance with the spin splitting, and the dipole $r_{\mathrm {dip}}$ between the orbital states vanishes in a harmonic confinement.]  
}
For both forms of SOC, $1/T_\varphi$ has similar dependence on the magnetic field $B_0$ and narrowly peaks at the SVH, which can be useful for system characterization.
Moreover, given that the spin pure dephasing from other mechanisms is at least $10^4$ $s^{-1}$ \cite{veldhorst2014, veldhorst2015, yoneda2018}, $1/T_\varphi$ due to the SVM is only relevant near the hot-spot. Therefore, slightly away from the hot-spot, before $1/T_\varphi$ due to the SVM starts to dominate spin dephasing, the vacuum Rabi frequency (and also the Rabi frequency of EDSR) could be enhanced by orders of magnitude by tuning the valley splitting $E_{\mathrm {VS}}$ or the magnetic field $B_0$ so that system is close to the point of SVH, while spin dephasing remains roughly constant.


\subsection{Quantum gate operation near the SVH}

%
%
The asynchronous rise of EDSR Rabi frequency $\Omega_{\mathrm R}$ and total spin dephasing near the SVH hints that one could possibly perform fast and high-fidelity quantum gates in this regime.  Considering that spin relaxation (as shown below) is generally slower than pure dephasing, even at the hot-spot, the total decoherence rate can be estimated as $1/T_2^* = 1/T_\varphi + 1/T_{\varphi,0}$, where $1/T_{\varphi,0}$ originates from other sources such as nuclear spins or charge noise via longitudinal gradient \cite{yoneda2018}. Recent experiments show that $1/T_{\varphi,0} \sim 5\times 10^4$ s$^{-1}$ for a spin qubit in an isotopically purified Si QD \cite{yoneda2018}. 
With $\Omega_{\mathrm R}$ characterizing how fast a single-qubit gate can be, the single-qubit quality factor can then be defined as $Q_{\mathrm{Rabi}}=\Omega_{\mathrm R} T_2^*/\pi$, which is a measure of how well one can control such a qubit.

\begin{figure}
\includegraphics[scale=0.6]{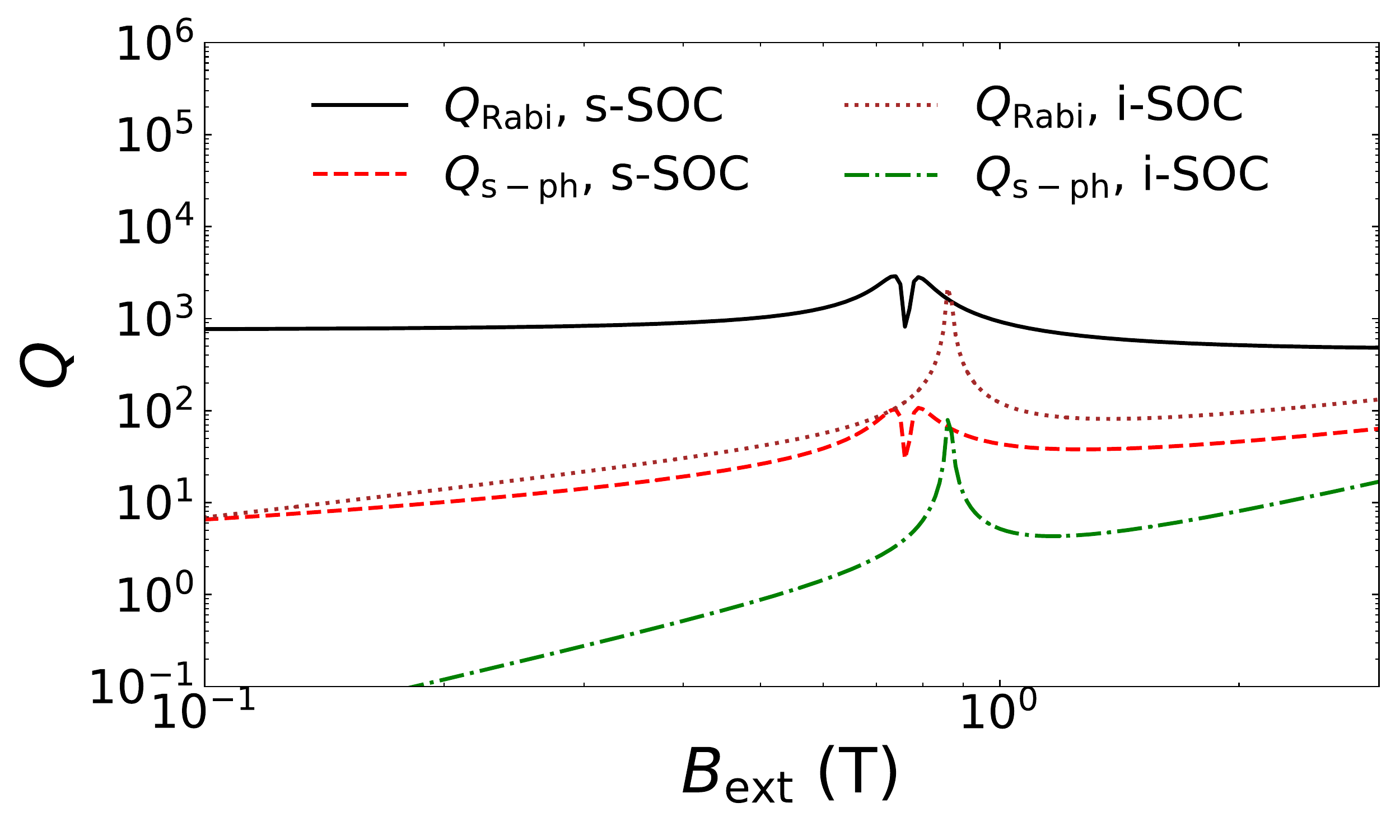}
\caption{\added{Quality factors for EDSR and spin-photon coupling.}
Quality factors for EDSR Rabi driving and spin-photon coupling as a function of the magnetic field $B_0$ for a spin in a silicon QD with the s-SOC or the i-SOC. 
A constant pure dephasing rate $5\times10^4$ $s^{-1}$ is assumed from other mechanisms \cite{yoneda2018}.
Sweet-spots are achieved near (not at) the SVH for the s-SOC, where the quality factors for the EDSR driving and spin-photon coupling are enhanced substantially. The quality factors are enhanced for the i-SOC as the system approaches the SVH and shows no dip at the SVH since the pure dephasing via SVM is not dominant near the SVH.
}\label{fig_Q}
\end{figure}

Fig. \ref{fig_Q} shows the quality factor $Q_{\mathrm{Rabi}}$ with the s-SOC or i-SOC as a function of the external magnetic field. $Q_{\mathrm{Rabi}}$ with the s-SOC exhibits two sweet-spots near (not at) the SVH, before dephasing due to the SVM starts to dominate. $Q_{\mathrm{Rabi}}$ increases by an order of magnitude as the system approaches the sweet-spots. If the spin qubit is coupled to a superconducting resonator, with the resonator decay rate $\kappa$ and the bare spin decoherence rate $1/T_{\varphi,0}$ both $\sim 5\times 10^4$ s$^{-1}$, the strong coupling limit of $g_{\mathrm s} > \kappa, 1/T_2^*$ can be achieved when the system approaches the SVH. Indeed, Fig.~\ref{fig_Q} shows that the quality factor $Q_{\mathrm{s-ph}}=g_{\mathrm s}T_2^*/(\pi)$ of spin-photon coupling can reach above $10^2$.  In short, by using the SVM near the hot-spot, fast high-fidelity quantum gates are within reach for Si quantum computing.

The trend for the qubit quality factor to reach its peaks near the SVH holds for the i-SOC as well.  Fig. \ref{fig_Q} also shows that the quality factors $Q_{\mathrm{Rabi}}$ and $Q_{\mathrm{s-ph}}$ without micromagnets can also be improved by orders of magnitudes near the SVH, which can be useful for high fidelity quantum gates without micromagnets, although the operation speed is slower than the case of the s-SOC.

{Besides decoherence, leakage error could occur due to transitions to states outside the computational basis. However, the leakage error is suppressed if the detuning between the level splitting and the driving field frequency is much larger than the bandwidth of the driving field. Note that the harmful processes that lead to the leakage error of a spin qubit are transitions involving simultaneous flipping of the spin and valley states [Suppose the spin dynamics is of interest and the valley degree of freedom is traced out, and the g-factor difference is assumed small between the valleys]. At the sweet spots, where $\epsilon_- = E_{\mathrm {VS}}-E_{\mathrm Z}$ is relatively small, we have $\epsilon_- \approx 0.03$ T $\sim$ 3 $\mu$eV, which is still much larger than the spin-valley coupling amplitude $\Delta$ of about 0.1 $\mu$eV. As such, the electron eigenstates are close to spin-valley product states.] Near the spin-valley sweet-spot, the estimated Rabi frequency is 
on the order of $2\pi\times10^7$ to $2\pi\times10^8$ 1/s, which corresponds to a B-field of 0.3 mT to 3 mT, much less than the detuning between the microwave and the level splitting (the levels coupled by the SOC and involves simultaneous spin-valley flip). Thus, the leakage should be largely suppressed by the energy detuning. Moreover, leakage error can further be suppressed by engineered pulses \cite{motzoi_simple_2009, chen_measuring_2016, sheldon_procedure_2016, werninghaus_leakage_2021}.
}


We emphasize that the valley splitting in Si can be tuned electrically using top gates \cite{yang2013, ibberson_electric-field_2018, zhang_giant_2020, hosseinkhani_electromagnetic_2020}. Thus, one can electrically tune the valley splitting to turn on the SVM for spin manipulation and spin-spin coupling, and turn off the SVM for the idling qubits. Furthermore, with the SVM boosted EDSR in a single QD instead of a double QD \cite{hu_strong_2012, mi_coherent_2018, benito_electric-field_2019, borjans_resonant_2020, croot_flopping-mode_2020}, a spin-qubit-based architecture could be simplified without sacrificing manipulation speed and tunability.

\subsection{Spin relaxation}

The SVM due to s-SOC means that electrical noises can cause spin relaxation. In particular, the resulting spin relaxation is
\beq
1/T_1 = \frac{4\pi e^2}{\hbar^2} \eta_{\mathrm {SV}}^2(\hbar\omega_{\mathrm Z}) \sum_i\left|{r}_i^{v_0v_1}\right|^2S_{ii, \mathrm {E}}(\omega_{\mathrm Z}), \label{T1}
\eeq
where $S_{ii, \mathrm {E}}(\omega)\equiv\frac{1}{2\pi}\int_{-\infty }^{+\infty }d\tau \overline{E_{i}(0)E_{i}(\tau)} \cos(\omega\tau)$ is the spectral density of noisy electric field ($i=x,y,z$).  When $E_{\mathrm Z}>E_{\mathrm {VS}}$, an additional spin relaxation channel via the intermediate state $\ket{\widetilde{2}}$ arises \cite{yang2013, huang_spin_2014}. However, its contribution to overall spin relaxation is relatively weak and is not included here (see Supplementary Note 1){, except for the relaxation due to the $1/f$ charge noise as shown below.}

An important feature of spin relaxation rate $1/T_1$ is its dependence on the magnetic field $B_0$. This $B_0$ dependence is determined by the spectral density $S_{ii, \mathrm {E}}(\omega_{\mathrm Z})$ of the noisy electrical field and the factor $\eta_{\mathrm {SV}}(\hbar\omega_{\mathrm Z})$ that captures the effect of the SVM.
Table \ref{Table_spinRelax} summarizes the $B_0$ dependence of $1/T_1$ via the SVM due to deformation potential (DP) phonon, piezoelectric (PE) phonon, Johnson noise, or $1/f$ charge noise at the zero temperature limit. At finite temperatures, an extra term $\coth(E_{\mathrm Z}/k_{\mathrm B}T)$ in spectral density of Johnson and phonon noise will play a role. In particular, at the high-temperature limit when $T>g\mu_{\mathrm B}B_0/k_{\mathrm B}$, $\coth(E_{\mathrm Z}/k_{\mathrm B}T) \approx k_{\mathrm B}T/(g\mu_{\mathrm B} B_0)$, which results in extra $1/B_0$ dependence for spin relaxation \cite{huang_spin_2014}.

In addition to spin relaxation due to the SVM, it could also arise due to intra-valley SOM, which can be obtained from the result of the SVM by replacing $E_{\mathrm {VS}}$ and $r_i^{v_0v_1}$ by $E_{\mathrm d}$ and $r_{\mathrm d}=r_{\mathrm {d0}}F_{\mathrm c}(\omega_{\mathrm Z})$, where $E_{\mathrm d}$ is the intra-valley orbital splitting, $r_{\mathrm {d0}}=\hbar/\sqrt{mE_{\mathrm d}}$ is the transition dipole between the lowest two orbitals, and $F_{\mathrm c}(B_0)=e^{-\omega_{\mathrm Z}^2r_{\mathrm {d0}}^2/(2v_j^2)}$ is due to the contribution beyond the electric dipole approximation and leads to phonon bottleneck effect at higher magnetic fields \cite{golovach2004, meunier_experimental_2007, huang_electron_2014}.



\begin{figure}
\includegraphics[scale=0.6]{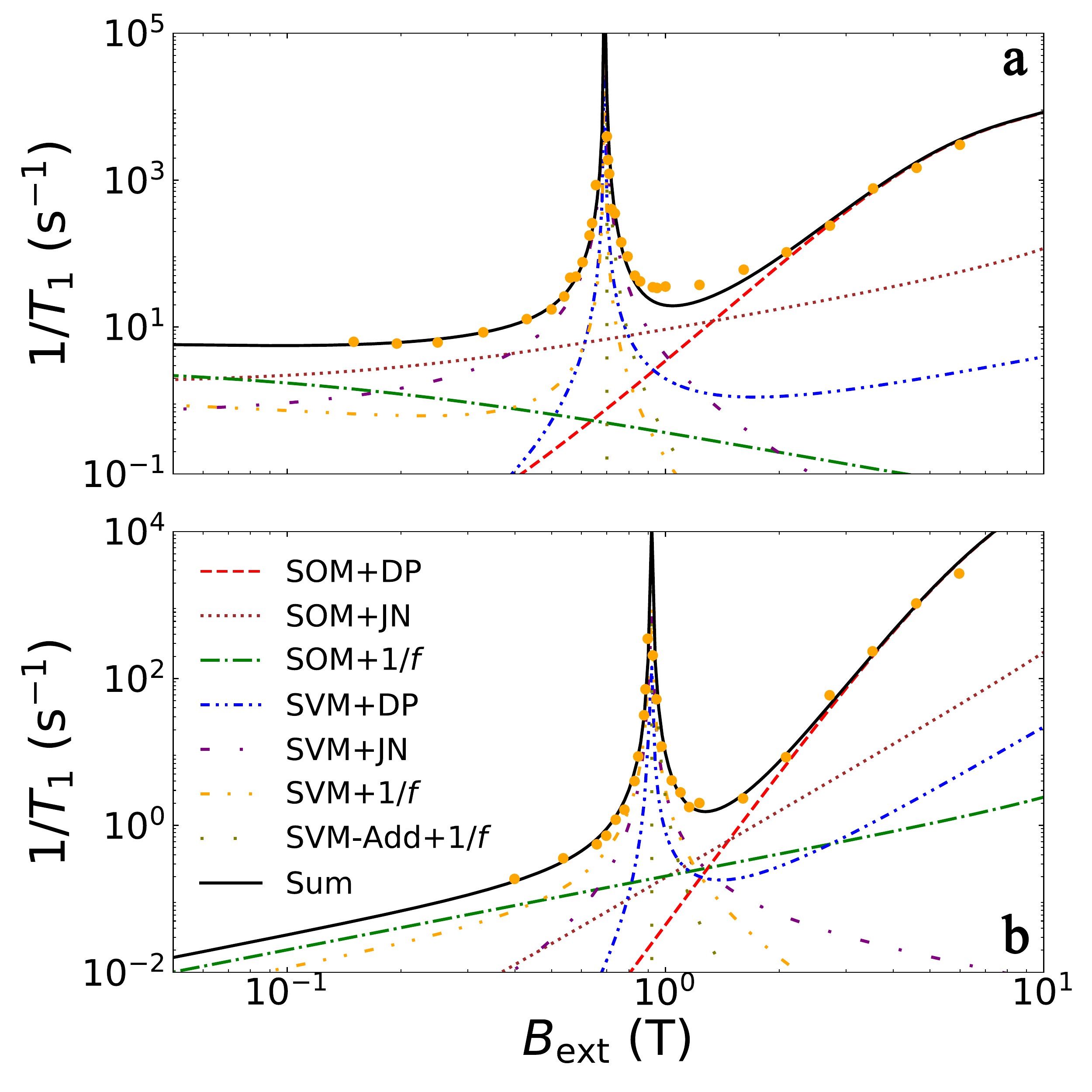}
\caption{\added{Spin relaxation versus magnetic field in silicon.} Spin relaxation $1/T_1$ in a silicon QD as a function of the magnetic field $B_0$ {due to the s-SOC (Fig. \ref{fig_spinRelax} a) or the i-SOC (Fig. \ref{fig_spinRelax} b)} and DP phonon, Johnson noise (JN), and $1/f$ charge noise via the SVM or the intra-valley SOM. 
\added{
When $E_{\mathrm Z} > E_{\mathrm {VS}}$, we also include the additional spin relaxation from $1/f$ charge noise (SVM-Add + $1/f$) via an intermediate state, whose magnitude is appreciable near the hotspot but fast decreases as $B_0$ goes away from hotspot.}
The dots are the experimental data from Ref. \cite{borjans_single-spin_2019}. 
{$1/T_1$ due to the s-SOC induced SVM (or intra-valley SOM)} shows weaker dependence with $B_0$ than the case of the i-SOC, which explains the saturation of $1/T_1$ at low magnetic fields and the $B_0$ dependence at high magnetic fields when micro-magnets present.
}\label{fig_spinRelax}
\end{figure}

Fig. \ref{fig_spinRelax} shows the spin relaxation via s-SOC (Fig. \ref{fig_spinRelax} a) or i-SOC (Fig. \ref{fig_spinRelax} b) for devices with or without a micromagnet.
The dots are the experimental data from Ref.~\cite{borjans_single-spin_2019}. The lines are theoretical results for DP phonon, Johnson noise, and $1/f$ charge noise via the SVM or the intra-valley SOM (parameters listed in Methods). Our theory faithfully captures the main features observed in the experiment. \added{In particular, the broken $T$-symmetry of the s-SOC leads to a weak $B_0$ dependence of spin relaxation compared with the case of the i-SOC as mentioned above.
With a micromagnet (Fig. \ref{fig_spinRelax} a), spin relaxation at low $B_0$ field saturates, and is dominated by the $1/f$ charge noise and Johnson noise. 
Specifically, $1/f$ charge noise via the s-SOC induced intra-valley SOM plays an important role in spin relaxation at very low $B_0$ field due to the large noise spectral density at low frequencies. As $B_0$ increases, spin relaxation becomes dominated by Johnson noise via the intra-valley SOM, and shows $B_0\coth(\gamma_eB_0/k_{\mathrm B}T_{\mathrm e})\sim k_{\mathrm B}T_{\mathrm e}/\gamma_e$ dependence that is independence of $B_0$ at the low field limit ($B_0\ll k_{\mathrm B}T_{\mathrm e}/(g\mu_{\mathrm B})$).}
In comparison, spin relaxation would not be saturated at low magnetic field in the case of the i-SOC, as shown in Fig. \ref{fig_spinRelax} b.

\added{As the uniform magnetic field $B_0$ further increases (near the SVH), spin relaxation becomes dominated by the SVM mechanisms and rises in a sharp peak, consistent with the experimental data.}
At high magnetic fields, when $E_{\mathrm Z}>E_{\mathrm {VS}}$, our theory again captures the main features of $1/T_1$. 
{Here, the additional spin relaxation (SVM-Add) via an intermediate state due to the $1/f$ charge noise is included when $E_{\mathrm Z}>E_{\mathrm {VS}}$ \added{[see Supplementary Note 1 for more information]}. Due to the small energy splitting, the $1/f$ charge noise can induce appreciable additional spin relaxation when $E_{\mathrm Z}\gtrsim E_{\mathrm {VS}}$. However, it fast decreases when $B_0$ is away from the hot-spot.}
At high magnetic fields, \added{the spin relaxation has two major contributions. One is due to the intra-valley SOM and Johnson noise that has a linear $B_0^3$ dependence.} The other is due to phonon noise via the intra-valley SOM, which has a $B_0^5$ dependence at first but is suppressed at higher $B_0$ due to the phonon bottleneck effect \cite{golovach2004, tahan2014, huang_spin_2014}. The suppression of spin relaxation at higher $B_0$ is indeed visible from the experimental data. 
\added{Consequently, the $B_0$-dependence for spin relaxation is always slower than $B_0^5$, and slow down even further at higher magnetic fields.}
Again notice that $1/T_1$ has a weaker $B_0$-dependence for the s-SOC compared to the case of the i-SOC, which is due to the different $T$-symmetries of the two SOC mechanisms. 
\deleted{Finally, near the SVH, $1/T_{1}$ rises in a sharp peak, consistent with the experimental data.}



\section{Discussion}
Our results suggest that the $T$-symmetry plays a vital role in the mixings of spin-valley states and determines spin properties in silicon. The fast EDSR at low magnetic due to s-SOC induced SVM makes the system compatible with superconducting circuits.
Furthermore, one can readily extend our results to different scenarios, such as an electron (or a hole) in a double quantum dot, where the results of spin dephasing hot-spot and spin manipulation sweet-spots can be applied to the case of mixed spin and charge states. 
Our study indicates that the mixing of spin states to valley states not only improves spin qubit control compared to a spin qubit well separated from the valley dynamics, but also save the number of QDs used for qubit encoding (i.e., a single dot rather than a double dot to confine a spin qubit).
It represents a great example that a hybrid quantum system offers an improvement over the individual quantum systems. 

In conclusion, we studied the theory of spin manipulation, pure dephasing, and relaxation due to SVM via the s-SOC from a micromagnet. We find 
the spin transition dipole mediated by the s-SOC {induced SVM (or intra-valley SOM) shows weak magnetic field dependence} arising from the broken $T$-symmetry of the s-SOC.
EDSR mediated by the s-SOC and SVM is enhanced as a result of the constructive interference and the large mixing at the SVH.
Furthermore, pure dephasing from the $1/f$ charge noise is possible due to SVM and s-SOC, and the SVH for relaxation is also a spin dephasing hot-spot.
Combining our results on Rabi frequency and spin dephasing, we find that the parameter regime near (but not at) the SVH may provide an optimal point for fast and high-fidelity quantum gates.
Our theory also explains the experimentally observed field dependence of spin relaxation at both low and high magnetic fields, which is not captured by previous theoretical results.
We hope our work will stimulate further explorations of the benefits of hybridized quantum systems, the valley degree of freedom, and the effects of symmetry and interference on solid-state qubits.

\section{Methods}

To study spin decoherence and manipulation in the system, we first obtain the spin-valley eigenstates in the presence of spin-valley coupling without environmental noises, and then evaluate the effective electric dipole matrix elements between the spin-valley mixed states. 
From the effective dipole moments and the potential from electrical noise and manipulation field, the spin relaxation, manipulation, and dephasing dynamics is evaluated.

\subsection{Spin-valley eigenstates and transition dipoles}
Consider the mixing between the states $|v_0\uparrow \rangle$ and $|v_1\downarrow \rangle$ due to the SOC, 
the detuning of the states $|v_0\uparrow \rangle$ and $|v_1\downarrow \rangle$ is $\varepsilon=E_{\mathrm {VS}}-E_{\mathrm Z}$. 
The coupling matrix element is $\Delta_{v_0\uparrow,v_1\downarrow}$. Since $\Delta_{v_0\uparrow,v_1\downarrow}$ is in general a complex number, we denote $\Delta_{v_0\uparrow,v_1\downarrow} = \Delta e^{\mathrm{i}\delta_-}$, where $\delta_-=\arctan[\IM(\Delta_{v_0\uparrow,v_1\downarrow})/\RE(\Delta_{v_0\uparrow,v_1\downarrow})]$ is the phase and $\Delta =|\Delta_{v_0\uparrow,v_1\downarrow}|$ is the magnitude of matrix element.
Then, the eigenstates in the subspace can be obtained
\beqa
\ket{\widetilde{3}}&=&\cos\frac{\gamma_-}{2}e^{-\mathrm{i}\delta_-/2} \ket{v_1\downarrow} + \sin\frac{\gamma_-}{2}e^{\mathrm{i}\delta_-/2}\ket{v_0\uparrow},\nonumber\\
\ket{\widetilde{2}}&=&-\sin\frac{\gamma_-}{2}e^{-\mathrm{i}\delta_-/2}\ket{v_1\downarrow}+\cos\frac{\gamma_-}{2}e^{\mathrm{i}\delta_-/2}\ket{v_0\uparrow},\nonumber
\eeqa
where $\gamma_-=\arctan(\Delta/\varepsilon)$. 
The energy splitting between the two eigenstates is $\widetilde{\varepsilon}=\sqrt{\varepsilon^2+\Delta^2}$. 

Similarly, there is also mixing between the states $|v_0\downarrow \rangle$ and $|v_1\uparrow \rangle$ due to the SOC. The detuning of the two states is $\varepsilon^\prime=E_{\mathrm {VS}}+E_{\mathrm Z}$, and coupling matrix element is $\Delta_{v_0\downarrow,v_1\uparrow} = \Delta^\prime e^{\mathrm{i} \delta_+}$, where $\Delta^\prime = |\Delta_{v_0\downarrow,v_1\uparrow}| =\Delta$, and  $\delta_+=\arctan[\IM(\Delta_{v_0\downarrow,v_1\uparrow})/\RE(\Delta_{v_0\downarrow,v_1\uparrow})]$. Then, the eigenstates in the subspace is obtained similarly with modified mixing angles $\gamma_+$ and $\delta_+$,
where $\gamma_+=\arctan(|\Delta_{v_0\downarrow,v_1\uparrow}|/\varepsilon^\prime)$. The energy splitting of the two eigenstates is $\widetilde{\varepsilon}^\prime=\sqrt{\varepsilon^{\prime 2}+\Delta^2}$.


%
According to the expression of the spin-valley eigenstates obtained above, the transition dipole matrix element between spin-valley eigenstates $\ket{\tilde{1}}$ and $\ket{\tilde{2}}$ is 
\beqa
\bra{\widetilde{1}}r_i\ket{\widetilde{2}}
&=&-|r_i^{v_0v_1}|\left[\cos \phi_i\sin\left(\frac{\gamma_-+\gamma_+}{2}\right)\right.  \nonumber \\
&& \left. + \mathrm{i}\sin \phi_i\sin\left(\frac{\gamma_--\gamma_+}{2}\right) \right],
\eeqa
where $r_i^{v_0v_1}=(r_i^{v_1v_0})^*\equiv \bra{v_0} r_i \ket{v_1} = |r_i^{v_0v_1}|e^{\mathrm{i}\phi_{r,i}}$ is the dipole matrix element between the valley states,
and $\phi_i = \phi_{r,i} - (\delta_-+\delta_+)/2$.
Similarly, from the expression of spin-valley eigenstates, the dipole matrix element between $\ket{\tilde{1}}$ and $\ket{\tilde{3}}$ can be obtained by changing $\sin(\gamma_-\pm\gamma_+)/2$ to $\cos(\gamma_-\pm\gamma_+)/2$ in the expression of $\bra{\widetilde{1}}r_i\ket{\widetilde{2}}$.

The dipole moment that is relevant to spin dephasing is $\bra{\widetilde{1}}\vec{r}\ket{\widetilde{1}}-\bra{\widetilde{2}}\vec{r}\ket{\widetilde{2}}$, in which
\beqa
\bra{\widetilde{2}} \vec{r} \ket{\widetilde{2}}
&=& \cos^2(\gamma_-/2)\vec{r}^{v_0v_0} + \sin^2(\gamma_-/2)\vec{r}^{v_1v_1},
\eeqa
\beqa
\bra{\widetilde{1}} \vec{r} \ket{\widetilde{1}}
&=& \cos^2(\gamma_+/2)\vec{r}^{v_0v_0} + \sin^2(\gamma_+/2)\vec{r}^{v_1v_1}.
\eeqa
Therefore, the dipole moment contributing to pure dephasing is 
\beq
\bra{\widetilde{1}}\vec{r}\ket{\widetilde{1}}-\bra{\widetilde{2}}\vec{r}\ket{\widetilde{2}} = \sin(\frac{\gamma_- +\gamma_+}{2})\sin(\frac{\gamma_- -\gamma_+}{2})\vec{r}_{\mathrm {dip}},
\eeq
which is proportional to $\vec{r}_{\mathrm {dip}}=(\vec{r}^{v_1v_1}-\vec{r}^{v_0v_0})$. Then, the noise contributing to spin dephasing is $n_{\mathrm {eff}}=e\vec{E}_{\mathrm {noise}} \cdot (\bra{\widetilde{1}}\vec{r}\ket{\widetilde{1}}-\bra{\widetilde{2}}\vec{r}\ket{\widetilde{2}})$. 
The $1/f$ charge noise is an importance source of pure dephasing. Suppose the noisy electric field from the $1/f$ charge noise on a single quantum dot is on the same order as the noise on a double quantum dot, and the noisy voltage fluctuation in a double quantum dot is $V_{1/f}$ (on the order of $1$ $\mu eV$), then, we have $|E_{\mathrm {noise}}|\sim V_{1/f}/l_0$, where $l_0$ is the typical length scale during the noise measurement on a double quantum dot.



\subsection{Broken $T$-symmetry of the s-SOC}
Now we prove that, for synthetic SOC, we have $\Delta_{v_0\uparrow,v_1\downarrow}^* =  \Delta_{v_0\downarrow,v_1\uparrow}$ for the mixing matrix element, so that $\delta_-+\delta_+=0$. For the dipole matrix element between the valley states, we find that $r_i^{v_0v_1}$ is real, $r_i^{v_0v_1}=r_i^{v_1v_0}$, so that $\phi_{r,i}=0$, and $\phi_i=\phi_{r,i} - (\delta_-+\delta_+)/2=0$. Note that the phases $\phi_{r,i}$ and $\delta_-+\delta_+$ actually depends on the convention of the global phase of the valley states, but the overall phase $\phi_i$ stay unchanged.

For the valley states in silicon, assuming the separability of the orbital and valley degree of freedom, the wave-function in the effective mass theory is $\langle{r}\ket{n}=\sum_j a_n^j F_n^j(\vec{r}) \psi_j(\vec{r})$, where $j$ runs from 1 to 6 for different valley states, and $n$ is index for the different orbital states.
In the case of a silicon QD, where the electron experiences a strong confinement in the vertical direction and an anisotropy of the effective mass, the valley states $x$, $\bar{x}$, $y$, $\bar{y}$ are much higher in energy than the $z$, $\bar{z}$ valley states. Then, we can consider only the $z$ and $\bar{z}$ valley states by neglecting the others. {The wave functions of the lowest two valley states are \cite{zhang_giant_2020}
$\ket{v_0}= \frac{1}{\sqrt{2}} \ket{F_0(\vec{r})} \left[e^{-\mathrm{i}k_0z}\ket{u_{-z}} + e^{\mathrm{i}\phi} e^{\mathrm{i}k_0z}\ket{u_{z}}\right]$, and
$\ket{v_1}= \frac{1}{\sqrt{2}} \ket{F_1(\vec{r})} \left[e^{-\mathrm{i}k_0z}\ket{u_{-z}} - e^{\mathrm{i}\phi} e^{\mathrm{i}k_0z}\ket{u_{z}}\right]e^{\mathrm{i}\phi_{v_1}}$,
where $F_{0}(\vec{r})$ and $F_{1}(\vec{r})$ are the envelope functions, and $\phi$ is the phase difference between the two valley states, and $\phi_{v_1}$ is the global phase of the state $\ket{v_1}$.}
Without loss of generality, we choose the global phase $\phi_{v_1}=\pi/2$. Next, we consider the properties of the time-reversal operator $\Theta$. 

{For the purpose of our discussion, the time-reversal operator is denoted as $\Theta = \sigma_y K$, where $K$ is the complex-conjugate operator that forms the complex conjugate of any coefficient that multiplies a ket (and stands on the right of $K$) \cite{sakurai_modern_1994, dresselhaus_group_2008}.} Under the time-reversal operation, the coordinate operator $\vec{r}$ is symmetric, i.e. $\Theta \vec{r} \Theta^{-1} =  \vec{r}$; While the spin operator $\vec{\sigma}$ is asymmetric, i.e. $\Theta \vec{\sigma} \Theta^{-1} = -\vec{\sigma}$. {For spin states, 
$\Theta \ket{\uparrow} = \mathrm{i} \ket{\downarrow}$, $\Theta \ket{\downarrow} = - \mathrm{i} \ket{\uparrow}$.
The valley states under the time-reversal is 
$\Theta \ket{v_0} = \frac{1}{\sqrt{2}} \ket{F_0(\vec{r})} \left[e^{\mathrm{i}k_0z}\ket{u_{z}} + e^{-\mathrm{i}\phi} e^{-\mathrm{i}k_0z}\ket{u_{-z}}\right] = e^{-\mathrm{i}\phi} \ket{v_0}$,
and $\Theta \ket{v_1} = - \frac{\mathrm{i}}{\sqrt{2}} \ket{F_1(\vec{r})} \left[e^{\mathrm{i}k_0z}\ket{u_{z}} - e^{-\mathrm{i}\phi} e^{-\mathrm{i}k_0z}\ket{u_{-z}}\right] = e^{-\mathrm{i}\phi} \ket{v_1}$.
\added{[Note that the envelope functions $\ket{F_0(\vec{r})}$ and $\ket{F_1(\vec{r})}$ are assumed to be real. This is justified by the fact that the cyclotron radius near spin-valley hotspot is much larger than the QD radius (especially in the growth direction considering that we have an in-plane field) so that magnetic field effect on electron orbitals can be neglected.]}
%
Since the s-SOC is asymmetric under time-reversal operation, i.e. $\Theta H_{\mathrm {s-SOC}} \Theta^{-1}= -H_{\mathrm {s-SOC}}$, the spin-valley mixing matrix element is}
\beqa
\Delta_{v_0\uparrow,v_1\downarrow} 
= \bra{\Theta (v_1 \downarrow) } \Theta H_{\mathrm {s-SOC}} \Theta^{-1} \ket{\Theta (v_0\uparrow)}
= \Delta_{v_1\uparrow, v_0\downarrow}. \nonumber 
\eeqa
Consequently, 
$\delta_-+\delta_+=0$. On the other hand, the dipole matrix element $\vec{r}^{v_0v_1}$ {satisfies the relation $\vec{r}^{v_0v_1} = \bra{\Theta (v_1) } \Theta \vec{r} \Theta^{-1} \ket{\Theta (v_0)} = \vec{r}^{v_1v_0}$, i.e. $r_i^{v_1v_0}$ is real and $\phi_{r,i}=0$.} Therefore, $\phi_i = \phi_{r,i} - (\delta_-+\delta_+)/2=0$, and the dipole matrix elements are given by
$\bra{\widetilde{1}}r_i\ket{\widetilde{2}} = -\vec{r}^{v_0v_1}\sin(\gamma_-/2+\gamma_+/2)$
and $\bra{\widetilde{1}}r_i\ket{\widetilde{3}} = \vec{r}^{v_0v_1}\cos(\gamma_-/2+\gamma_+/2)$.
When $\gamma_-\gg\gamma_+$, these matrix elements take the approximate expressions of $\bra{\widetilde{1}}r_i\ket{\widetilde{2}}\approx -\vec{r}^{v_0v_1}\sin(\gamma_-/2)$ and $\bra{\widetilde{1}}r_i\ket{\widetilde{3}}\approx \vec{r}^{v_0v_1}\cos(\gamma_-/2)$, consistent with the results in our previous study \cite{huang_spin_2014}.

In comparison, for the case of the intrinsic SOC, $\Delta_{v_0\uparrow,v_1\downarrow} = -\Delta_{v_0\downarrow,v_1\uparrow}^*$,
which means that $\phi_i = \phi_{r,i} - (\delta_-+\delta_+)/2=\pi/2$. Thus, the sign in front of $\gamma_+$ changes in $\bra{\widetilde{1}}r_i\ket{\widetilde{2}}$ and $\bra{\widetilde{1}}r_i\ket{\widetilde{3}}$. 

{
\subsection{SOC matrix elements}
When the $x$ and $y$ axes are defined along the $[100]$ and $[010]$ crystallographic directions, the i-SOC is $H_{\mathrm {i-SOC}}=H_{\mathrm R}+H_{\mathrm D}$, where
$H_{\mathrm R} = \alpha_{\mathrm R} (p_x\sigma_y - p_y \sigma_x)$ 
and $H_{\mathrm D} = \alpha_{\mathrm D} (-p_x\sigma_x + p_y \sigma_y)$
are the Rashba and the Desselhaus SOC due to structural inversion asymmetry (SIA) and bulk inversion asymmetry (BIA), and $\alpha_{\mathrm R}$ and $\alpha_{\mathrm D}$ are the coupling constants \cite{golovach2004}. There is no BIA in bulk silicon. However, a Dresselhaus-like term can appear when an electron is near an interface \cite{prada_spinorbit_2011, ferdous_valley_2018, jock_silicon_2018, tanttu_controlling_2019}. In the following, we consider both the contributions of Rashba and Desselhaus SOC to the mixing of the spin-orbital states. 

When the coordinate is redefined so that the $x$ and $y$-axes are along the $[110]$ and $[\bar{1}10]$ directions, the SOC Hamiltonian is rewritten as \cite{golovach2004}
\beqa
H_{\mathrm {i-SOC}} &=& \alpha_- p_y \sigma_x + \alpha_+ p_x \sigma_y,
\eeqa
where $\alpha_\pm = \alpha_{\mathrm D} \pm \alpha_{\mathrm R}$.
In the experiment reported in Ref.~\cite{borjans_single-spin_2019}, the magnetic field is along the $[110]$ direction. As such the $\sigma_y$ term is transverse to $\vec{B}_0$ and leads to the mixing of spin-orbit states. Thus we consider the orbital states $\ket{0_x}$, $\ket{1_x}$ due to the confinement along the $[110]$ direction. The intra-valley spin-orbit coupling matrix element $\Delta_{0\uparrow,1\downarrow} \equiv \bra{0_x \uparrow} H_{\mathrm {i-SOC}} \ket{1_x \downarrow}$ due to the i-SOC is thus
\beqa
\Delta_{0\uparrow,1\downarrow} &=& \alpha_+ \bra{0_x}p_x\ket{1_x} \bra{\uparrow} \sigma_y \ket{\downarrow} 
= - E_{\mathrm d} x_{01}/\lambda_{so},
\eeqa
where $x_{01}=\hbar/\sqrt{m^*E_{\mathrm d}}$ and $\lambda_{so}=\hbar/(m^*\alpha_+)$ is the spin-orbit length. Similarly, $\Delta_{0\downarrow,1\uparrow} \equiv  \bra{0_x \downarrow} H_{\mathrm {i-SOC}} \ket{1_x \uparrow} = E_{\mathrm d}x_{01}/\lambda_{so} = - \Delta_{v_0\uparrow,v_1\downarrow}$, which exhibits the same relation as the spin-valley coupling matrix elements [the relation obtained based on the $T$-symmetry].

The s-SOC is $H_{\mathrm {s-SOC}} = \frac{1}{2}g\mu_{\mathrm B} b_{\mathrm {1t}} \sigma_z x$, 
where $b_{\mathrm {1t}}\equiv \partial B_z/\partial x$ is the transverse magnetic field gradient along the $x$ axis, i.e. [110]. The coupling matrix element $\Delta_{0\uparrow,1\downarrow} \equiv \bra{0_x \uparrow} H_{\mathrm {s-SOC}} \ket{1_x \downarrow}$ due to the s-SOC is then
\beq
\Delta_{0\uparrow,1\downarrow} = \beta_{\mathrm {1t}}\bra{0_x}x\ket{1_x} \bra{\uparrow} \sigma_z \ket{\downarrow}=\beta_{\mathrm {1t}} x_{01},
\eeq
where $\beta_{\mathrm {1t}}\equiv \frac{1}{2}g\mu_{\mathrm B} b_{\mathrm {1t}}$. Similarly, $\Delta_{0\downarrow,1\uparrow} \equiv  \bra{0_x \downarrow} H_{\mathrm {s-SOC}} \ket{1_x \uparrow} = \beta_{\mathrm {1t}} x_{01} =  \Delta_{0\uparrow,1\downarrow}$, which exhibits the same relation as the spin-valley coupling.
}

{
\subsection{Noise model}
The deformation phonon Hamiltonian has been studied in the literature \cite{golovach2004, yang2013, tahan2014, huang_spin_2018}.
For electrical noises, such as Johnson noise or $1/f$ charge noise, the photon wave vector is much larger than the size of a QD. The noise Hamiltonian can thus be expressed in the limit of dipole approximation as
\beq
V_{\mathrm e}(\vec{r})=-\vec{r}\cdot \vec{F}(t),
\eeq
where $\vec{F}(t)$ is the electric field due to a given electrical noise.

Here we give explicitly the power spectral densities of the electrical noises that give rise to spin relaxation and dephasing in the system. Suppose the noise of the circuits outside the dilution refrigerator is well-filtered. Johnson noise should then be mostly due to the low-temperature circuit (such as the metallic gate on top of the QD) inside the dilution refrigerator. The corresponding spectral density $S_{ii, \mathrm E}$ of electrical field is 
$S_{ii, \mathrm E} \left( \omega \right) = S_{\mathrm {V,JN}} (\omega)/(el_{0})^{2}$, 
where $i=X,Y$, or $Z$ directions, $S_{\mathrm {V,JN}}(\omega)$ is the spectral density of the voltage fluctuation due to Johnson noise, $e$ is the electron charge, and $l_{0}$ is the distance between the gate and the QD that converts the voltage fluctuations to the fluctuations of the electric field at the dot. The spectral density $S_{\mathrm {V,JN}}(\omega)=\frac{1}{2\pi }\int_{-\infty }^{+\infty }\overline{V\left( 0\right)  V\left( t\right) } \cos \left( \omega t\right) dt$ of the voltage fluctuations due to Johnson noise is \cite{huang_spin_2014} 
\begin{equation}
S_{\mathrm {V,JN}}\left( \omega \right) =2\xi \omega \hbar ^{2}f_{\mathrm c}(\omega_{\mathrm Z})\coth \left( \hbar \omega /2k_{\mathrm B}T\right), \label{Sv_John}
\end{equation}
where $\xi =R/R_{k}$ is a dimensionless constant, $R_{k}=h/e^{2}=26$ k$\Omega $ is the quantum resistance, and $R$ is the resistance of the circuit. $f_{\mathrm c}(\omega)=1/[1+\left( \omega/\omega _{R}\right) ^{2}]$ is a natural cutoff function for Johnson noise, where $\omega _{R}=1/RC$ is the cutoff frequency, and $C$ is the equivalent capacitor in parallel with the resistance $R$.


Another electrical noise that is ubiquitous in solid state material is the $1/f$ charge noise.  The typical noise spectral density is 
\beqa
S_{V,1/f}=A/\omega,
\eeqa
where $\sqrt{A}$ is the strength of the noise. We assume that the charge noise is from the slow fluctuators in dielectric material near the metallic gates. The corresponding spectrum of electric field is estimated as $S_{ii, \mathrm {E}} \left( \omega \right) = S_{V,1/f} (\omega)/(el_{0})^{2}$, 
where $l_0$ is the distance between the dielectric material and the QD.
}

\subsection{Physical parameters}
The following values of parameters are used if not specified. We choose g = 2, $m^* = 0.19m_0$, and $E_{\mathrm d} = 3$ meV for the effective g-factor, the effective mass, and the horizontal orbital confinement of an electron in a silicon QD.
{The transition dipole moments between valley states are $r_x=r_y=r_z=1$ nm, and the dipole moment $r_{i}^{v_0v_0}-r_{i}^{v_1v_1}=1$ nm.} 
The valley splitting is $E_{\mathrm {VS}}=0.1$ meV.
The vertical confinement length of the QD is $d_z=5$ nm.
%
For the SOC constants, we choose the Rashba constant as {$\alpha_{\mathrm R}=20$ m/s, and the Dresselhaus constant $\alpha_{\mathrm D}=80$ m/s} for rough estimation.
We choose {$b_{\mathrm {1t}}=1.8$ mT/nm} for the magnetic field gradient \cite{borjans_single-spin_2019}. 
A magnetic field $B\mu_{\mathrm B}=0.1$ T is assumed from a fully polarized micromagnet \cite{borjans_single-spin_2019} so that the total magnetic field $B_0=B_{\mathrm {ext}}+B_{\mathrm {MM}}$, where $B_{\mathrm {ext}}$ is the externally applied magnetic field.

For the evaluation of EDSR Rabi frequency, we choose $E_0=10000$ V/m for the maximum electric field of the microwave driving. While for spin-photon coupling, we use $E_{\mathrm {zpf}}=V_{\mathrm {zpf}}/l_0=\omega_0\sqrt{\hbar Z_0}/l_0$, where $V_{\mathrm {zpf}}$ is the voltage due to the zero-point fluctuation (ZPF) in the resonator, $\omega_0$ and $Z_0$ the frequency and the characteristic impedance of the resonator, and $l_0$ the length for the voltage drop \cite{blais_cavity_2004, hu_strong_2012}. We choose resonator impedance $Z_0=50$ $\Omega$, and the resonator frequency the same as the spin Larmor frequency.

For $1/f$ charge noise, we choose the noise amplitude {$\sqrt{A}=3$ $\mu eV/\sqrt{\text{Hz}}$, and the length scale $l_0=30$ nm inspired by the geometry and size of the quantum dot [assuming the source of noise is distributed in the dielectric material near the QD] \cite{borjans_single-spin_2019}.} 
For phonon noise, we choose $v_1=5900$ m/s and $v_2=v_3=3750$ m/s for the speed of the different acoustic phonon branches, $\rho_c=2200$ kg/m$^3$ for the mass density,
$\Xi_d=5.0$ eV and $\Xi_u=8.77$ eV for the dilation and shear deformation potential constants \cite{tahan2014,huang_spin_2014}.
The phonon temperature is set to be zero for simplicity. 

Next, we give explicitly the values of the physical parameters during the fitting of spin relaxation results.

For the fitting of the s-SOC results \cite{borjans_single-spin_2019}, we choose the valley splitting {$E_{\mathrm {VS}}=0.096087$ meV (equal to the Zeeman energy at $0.83$ T)}. {$B_{\mathrm {MM}}=0.14$ T} is assumed from a fully polarized micromagnet (if a different $B_{\mathrm {MM}}$ field was chosen, the data of spin relaxation can be fitted equally well, but the other fitting parameters will be slightly modified). We choose the orbital splitting \added{$E_{\mathrm d}=2.8$ meV} [The magnetic field gradient is finite only when the electron is moving along the $x$-axis. For the spin relaxation (or EDSR) due to the s-SOC induced spin-orbit mixing, the relevant confinement is along the $x$-axis, and can be chosen differently from the value $2$ meV of the orbital confinement reported in the experiment. Here $E_{\mathrm d}$ is the only adjustable parameter to fit the spin relaxation at high magnetic fields, if the phonon parameters and  $b_{\mathrm {1t}}=1.8$ mT/nm (from the experimental estimation \cite{borjans_single-spin_2019}) are kept fixed], the valley transition dipole moment \added{$r_x^{v_0v_1}=1.3$ nm}, and the dipole moment \added{$|r_i^{v_0v_0}-r_i^{v_1v_1}|=0.8$ nm}. Resistance for Johnson noise \added{$R_{\mathrm {JN}}=3.0$ $\Omega$}, electron temperature of Johnson noise {$T_{\mathrm e}=115$ mK}. Charge noise amplitude \added{$\sqrt{A}=1$ $\mu$eV/$\sqrt{\text{Hz}}$}, low cutoff frequency $\omega_{c0}=1$ Hz, {and the length $l_0=30$ nm, which is inspired by the geometry and size of the quantum dot}. 

For the fitting of the i-SOC results, we choose {the orbital splitting $E_{\mathrm d}=3.9$ meV}, the valley splitting $E_{\mathrm {VS}}=0.105927$ meV (equivalent to a Zeeman energy at $0.915$ T), valley transition dipole moment \added{$r_x^{v_0v_1}=1.1$ nm}, and the dipole moment \added{$|r_i^{v_0v_0}-r_i^{v_1v_1}|=0.3$ nm}. Resistance for Johnson noise \added{$R_{\mathrm {JN}}=5$ $\Omega$}, electron temperature of Johnson noise $T_{\mathrm e}=115$ mK; Charge noise amplitude \added{$\sqrt{A}=4.5$ $\mu$eV/$\sqrt{\text{Hz}}$}, the length $l_0=30$ nm, low cutoff frequency $\omega_{c0}=1$ Hz; Rashba and Desselhaus SOC coupling constants are \added{$\alpha_{\mathrm R}=50$ m/s and $\alpha_{\mathrm D}=280$ m/s}.

\section{Data availability}

The main data supporting the finding of this study
are available within the article and its Supplementary Information files.
Additional data can be provided upon request.

\section{Acknowledgments}
The authors thank J. R. Petta and F. Borjans for helpful discussions.
P.H. acknowledges supports by the National Natural Science Foundation of China (No. 11904157), Shenzhen Science and Technology Program (No. KQTD20200820113010023), and Guangdong Provincial Key Laboratory (No. 2019B121203002); X.H. acknowledges support by US ARO via grant W911NF1710257.
\section{Competing interests} The authors declare no financial or non-financial conflicts of interest.

\section{Author contributions}
P.H. performed derivation and numerical calculation. P.H. and X.H. researched, analyzed, and prepared the manuscript.





\begin{table}
\begin{tabular}{|c|l|l|l|l|l|l|}
  \hline
   $1/T_1$ & DP & PE & JN & $1/f$ & general case \\
  \hline
  i-SOC ($E_{\mathrm Z}<E_{\mathrm {VS}}$) & $B_0^7$ & $B_0^5$ & $B_0^3$ & $B_0$ & $\propto \omega_{\mathrm Z}^2S(\omega_{\mathrm Z})$ \\
  i-SOC ($E_{\mathrm Z}>E_{\mathrm {VS}}$) & $B_0^3$ & $B_0$ & $B_0^{-1}$ & $B_0^{-3}$ & $\propto \omega_{\mathrm Z}^{-2}S(\omega_{\mathrm Z})$  \\
  s-SOC ($E_{\mathrm Z}<E_{\mathrm {VS}}$) & $B_0^5$ & $B_0^3$ & $B_0$ & $B_0^{-1}$ & $\propto S(\omega_{\mathrm Z})$ \\
  s-SOC ($E_{\mathrm Z}>E_{\mathrm {VS}}$) & $B_0$ & $B_0^{-1}$ & $B_0^{-3}$ & $B_0^{-5}$ & $\propto \omega_{\mathrm Z}^{-4}S(\omega_{\mathrm Z})$ \\
\hline
\end{tabular}
\caption{\added{Magnetic field dependences of spin relaxation for different noises and SOC.} Spin relaxation rate $1/T_1$ versus the magnetic field $B_0$ due to DP phonon, piezo-electric (PE) phonon, Johnon noise (JN), or $1/f$ charge noise via the i-SOC or the s-SOC (at zero temperature limit). $S(\omega_{\mathrm Z})$ is the power spectral density of electric field, $\omega_{\mathrm Z}=g\mu_{\mathrm B} B_0/\hbar$ is the Larmor frequency.
}\label{Table_spinRelax}
\end{table}

\clearpage


%
%
%
%
%
%
%

\end{document}